\documentclass[journal]{IEEEtran}

\usepackage{booktabs}
\usepackage{threeparttable}
\usepackage{amsmath,graphicx,cite,amssymb}
\usepackage{amsfonts,multirow,bm,array,setspace,stfloats}

\usepackage{graphicx,float,cite,amssymb}
\usepackage{graphics}
\DeclareGraphicsExtensions{.eps,.jpeg,.png,.jpg}   
\usepackage{multirow,bm,bbm,array,setspace}
\usepackage{textcomp}
\usepackage{psfrag}
\usepackage{color}
\usepackage{pstricks,enumerate}
\usepackage{bbm}
\usepackage{makecell}
\usepackage{xpatch}
\usepackage{subfigure}
\usepackage{enumitem}
\usepackage{algorithm,algpseudocode}
\usepackage{xpatch}
\makeatletter
\newcommand{\distas}[1]{\mathbin{\overset{#1}{\kern\z@\sim}}}%
\newsavebox{\mybox}\newsavebox{\mysim}
\newcommand{\distras}[1]{%
	\savebox{\mybox}{\hbox{\kern1pt$\scriptstyle#1$\kern1pt}}%
	\savebox{\mysim}{\hbox{$\sim$}}%
	\mathbin{\overset{#1}{\kern\z@\resizebox{\wd\mybox}{\ht\mysim}{$\sim$}}}%
}

\ExplSyntaxOn
\cs_new:Npn \bibColoredItems #1#2
  {
    \clist_map_inline:nn {#2} { \cs_new:cpn {bib@colored@##1} {#1} } 
  }
\ExplSyntaxOff

\newcommand\bib@setcolor[1]{%
  \ifcsname bib@colored@#1\endcsname
    \expandafter\color\expandafter{\csname bib@colored@#1\endcsname}
  \else
    \normalcolor
  \fi
}

\xpatchcmd\@bibitem
  {\item}
  {\bib@setcolor{#1}\item}
  {}{\PatchFailed}

\xpatchcmd\@lbibitem
  {\item}
  {\bib@setcolor{#2}\item}
  {}{\PatchFailed}

\makeatother
\ifCLASSINFOpdf
\else
\fi

\allowdisplaybreaks[4]

\begin{document}
	%
\title{Eliminating Blind Spots from Wireless Network by Metasurface: A Blind Approach}
\author{
\IEEEauthorblockN{
    Wenhai Lai, Mingxiao Li, Kaiming Shen,  Liyao Xiang, and Zhi-Quan Luo
}
\thanks{
    Accepted to IEEE Communications Magazine on 23 May 2026.
    Wenhai Lai is with Dalian University of Technology; Mingxiao Li, Kaiming Shen (corresponding author), and Zhi-Quan Luo are with The Chinese University of Hong Kong (Shenzhen) and Shenzhen Research Institute of Big Data; Liyao Xiang is with Shanghai Jiao Tong University.
    
}
}

%


\maketitle

\begin{abstract}
Deploying metasurfaces (MTSs) to eliminate wireless blind spots requires jointly determining the physical placement of MTSs and the meta-atom phase shifts. Existing methods typically rely on explicit channel estimation, which incurs prohibitive overhead and is often intractable in real-world networks. To sidestep this bottleneck, we propose RFZero, a channel-state-information (CSI)-free deployment paradigm. Instead of estimating channels, RFZero extracts macro-environmental features from visual photos to guide MTS placement, and leverages reference signal received power (RSRP) feedback for dynamic phase-shift optimization. Most importantly, RFZero operates independently of base stations, thereby enabling seamless plug-and-play implementation. Real-world field tests confirm that RFZero completely eliminates all blind spots in a $100\text{ m}^2$ indoor area using just a pair of $1.5\text{ m}\times 0.9\text{ m}$ MTSs.
\end{abstract}


\section{Introduction}
\label{sec:intro}

Metasurface (MTS) is formed by an array of meta-atoms, each inducing a tunable phase shift into the reflected propagation path. A common application of MTS is to eliminate blind spots from the wireless network for the signal coverage improvement. The following three technical questions are at the core of the MTS deployment: 
\begin{enumerate}[label=\emph{Q\arabic*:},leftmargin=*]
    \item How many MTSs are needed?
    \item Where are these MTSs placed?
    \item How to decide phase shift for each meta-atom of MTSs? 
\end{enumerate}
The existing methods in the literature are mostly based on a two-stage paradigm: first estimate channels so as to formulate the above questions as explicit optimization problems, and then resort to the standard methods. 

Channel acquisition, however, would impose a formidable challenge in practice, for the following reasons:
\begin{enumerate}[leftmargin=*]
\item \emph{Estimation Error:} It is fairly difficult to get precise estimation of the reflected cascaded channels induced by the MTSs, because each reflected path alone can be easily overwhelmed by the background noise plus interference.
\item \emph{Overhead Cost:} The pilot overhead for channel estimation would be quite costly if the meta-atoms are massively deployed at each MTS, e.g., the MTS prototypes used in our field tests have almost 300 meta-atoms each.
\item \emph{Protocol Compatibility:} Channel estimation typically requires multi-party cooperation among transmitter, receiver, and MTS, so it may not be compatible with the current network protocol. The implementation becomes particularly hard when it involves base stations.
\end{enumerate}
As such, it turns out that the existing prototype systems \cite{Arun2020RFocus,pei2021ris} seldom consider channel estimation for MTS.

\begin{figure*}[t]
    \centering
\includegraphics[width=\linewidth]{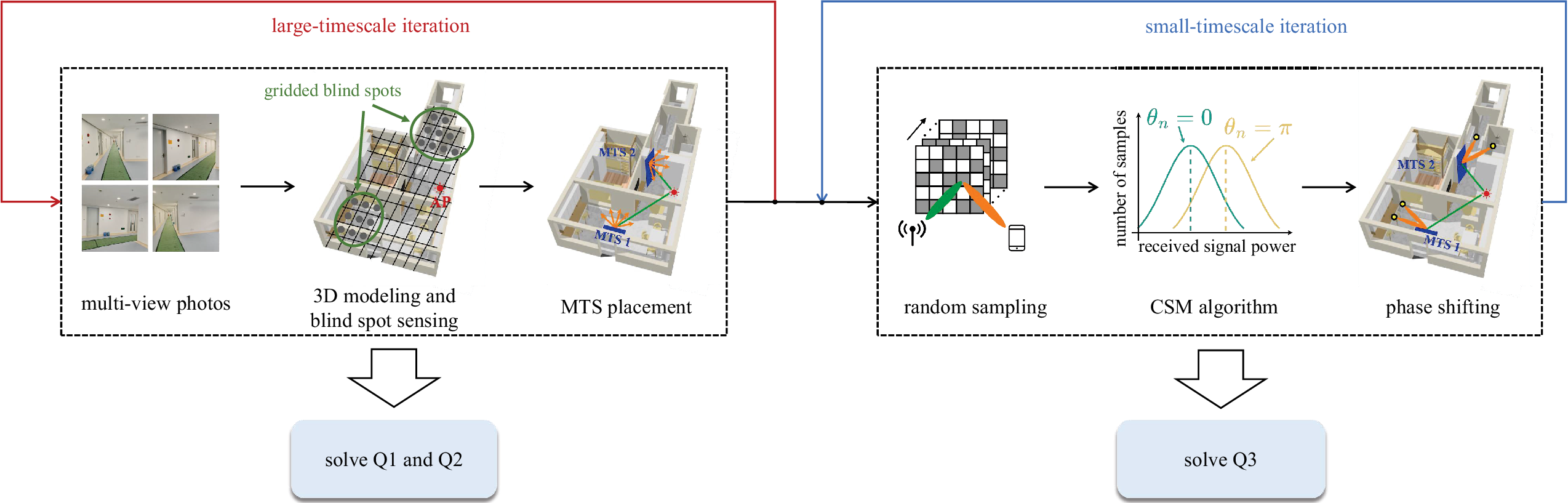}
    \caption{Overview of the proposed blind approach called RFZero for the MTS deployment. First, with the visual camera data (multi-view photos or a panoramic short video) as input, RFZero detects the blind spot clusters and places MTSs accordingly; second, with the received signal power data as input, RFZero optimizes phase shifts for the deployed MTSs so that the reflection beams are oriented to the target receivers.}
    \label{fig:RFZero}
\end{figure*}

To avoid these issues, our work proposes a channel-state-information (CSI)-free strategy called RFZero---which indicates deploying MTS to enhance the radio frequency (RF) environment with zero prior knowledge of channels. The proposed RFZero aims to extract the key knowledge of wireless environment from the camera data and the RSS measurements. The advantage of RFZero is that it avoids explicit instantaneous CSI estimation. In other words, RFZero only learns the ``useful'' knowledge of channels for the MTS placement and phase optimization, rather than the full knowledge of CSI. Importantly, to make it practical, the above questions Q1--Q3 are dealt with in two different timescales. The number of MTSs and where they are placed, as considered in Q1 and Q2, are decided in the long run, i.e., we update the MTS deployment only in response to the long-term environmental dynamics. In contrast, the phase shifting of meta-atoms, as considered in Q3, is performed much more frequently, e.g., every hour.

For Q1 and Q2, RFZero aims to extract the geographic features of the wireless network from the visual camera data (e.g., multi-view photos or a panoramic short video), based on which the (gridded) blind spots can be sensed via ray tracing. With the distribution of blind spots available, we further use geometry to decide the number of MTSs and their placements. Regarding Q3, we propose a statistical learning method. Analogically speaking, suppose that we wish to solve an optimization problem $\max f(x)$ but that the objective function $f(x)$ is unknown. As opposed to the conventional approach \cite{zheng2019intelligent} that seeks to recover $f(x)$, our method just tries out a series of random inputs $x$ and uses the output statistics to find the solution. To be more specific, RFZero decides the phase shifts based on the \emph{conditional sample mean} (CSM) of the reference signal received power (RSRP). While the performance guarantee of this CSM algorithm has been shown in the recent work \cite{xu2024blind}, the novelty of this paper lies in the integration of the CSM algorithm with the aforementioned visual data-based algorithm, which leads us to a blind approach to Q1--Q3 without requiring any channel estimation, as illustrated in Fig.~\ref{fig:RFZero}. The core challenge lies in bridging two entirely different physical domains and timescales: visual data represents macro-scale geometric optics (which changes slowly), while the CSM algorithm handles micro-scale electromagnetic fading (which fluctuates rapidly). Integrating them is not a simple concatenation; rather, it requires a novel dual-timescale decoupling architecture. Moreover, note that RFZero does not involve base stations, so it has a plug-and-play implementation.

\section{Related Work}


The previous works \cite{2024GPMS,PMSat2023} both adopt a free space channel model wherein the channel magnitudes and phases purely depend on the distance, so all the channels in this model can be precisely predicted given the device positions. As a result, the MTS placement boils down to an optimization problem of the position coordinates of MTS. To address the nonconvexity of this problem, \cite{2024GPMS,PMSat2023} suggest the convex approximation. When the MTS placement is restricted to a discrete set of possible spots, the method in \cite{efrem2023joint} first relaxes the position to be continuous and then rounds the continuous solution to the discrete constraint set. The above works all heavily depend on the distance-dependent free-space RF propagation model, assuming line-of-sight (LOS) propagation between any two spots within the geographic area of the wireless network. In contrast, the authors of \cite{AutoMS2024} propose a scheme called AutoMS that does not make any particular assumptions about the RF environment. AutoMS consists of two steps: first, it performs the 3D scanning to estimate the channels; second, with the channel estimate, it optimizes the MTS position by the heuristic algorithm of simulated annealing.

In comparison, only a few works in the existing literature account for Q1. For instance, \cite{efrem2023joint} only considers optimizing the number of MTSs as a byproduct of the MTS placement. Thus, \cite{efrem2023joint} still uses the relaxation method to decide the number of MTSs. Again, these methods are designed under the free space channel model. We remark that the above works only verify their methods in simulations. To the best of our knowledge, RFZero is the first algorithm that has been implemented in a real-world network to optimize the number of MTSs.

Many existing works assume that the CSI is perfectly known \emph{a priori} for the phase shifting task. When the phase shift is allowed to be chosen arbitrarily for each meta-atom, the resulting continuous problem is most relevant to semi-definite programming (SDR) and fractional programming (FP). When the phase shifts are limited to a discrete set as encountered in most practical situations, the authors of \cite{mu2021capacity} suggest exhaustive search for the global optimum, but it incurs exponential running time. Actually, even for a modest MTS containing $50$ meta-atoms and providing two phase shift options $\{0,\pi\}$ for each meta-atom, there are $2^{50}\approx 10^{15}$ possible solutions to explore, so the exhaustive search cannot finish within reasonable time. To get rid of the curse of dimensionality, a natural idea from \cite{RFBouncer2023} is to first relax the discrete constraint and then round the continuous solution to the discrete set, while \cite{SmartShell2023} proposes several heuristic methods including the simulated annealing and the genetic algorithm. 
Channel estimation is critical to these phase shifting algorithms because they all assume that the CSI is perfectly available.

\section{Long-Run Configuration}
\label{sec:MTS placement}

The long-run configuration aims to eliminate all the blind spots from the target network by deploying the smallest possible number of MTSs in the absence of CSI. 


\subsection{3D Modeling of Wireless Environment}
\label{subsec:3D_modeling}

Our approach is outlined in the following: we first perform the 3D modeling of wireless environment based on the visual camera data, then produce a wireless heatmap by ray tracing, then further sense and cluster blind spots, and ultimately decide how many MTSs are used and where they are placed according to geographic geometry.

We begin with the task of 3D modeling. Any modern smartphone equipped with visual scanning sensors (e.g., LiDAR) is sufficient for the task. Through the API \emph{RoomPlan}, an existing application software named \emph{3D scanner} is capable of identifying those common objects including door, wall, table, and chair, thereby allowing us to construct the 3D model of the target network scenario.

The 3D model inaccuracies typically stem from incomplete/blurry photos or incorrect electromagnetic parameters. To address this, RFZero employs a multi-tiered tolerance mechanism. First, a screening process mitigates input errors by warning the operator of low-quality images. Importantly, at the algorithmic level, RFZero is intentionally designed to be tolerant of model inaccuracies. The visual 3D model and ray tracing are only tasked with identifying the approximate macro-locations of blind-spot clusters (Q1 \& Q2). We do not rely on the 3D model to be perfectly accurate because the subsequent short-term CSM phase optimization (Q3) acts as an active error-absorption mechanism. If the MTS placement is slightly sub-optimal due to a 3D model error, the CSM algorithm empirically ``searches'' and aligns the real-world multipath reflections to compensate for this geometric deviation. Furthermore, as illustrated in Fig. \ref{fig:RFZero}, RFZero is a closed-loop system; if the system feedback shows that blind spots persist, engineers can quickly fine-tune the deployment.

We then discuss the practical advantages of our approach. Regarding the difficulty of acquisition, capturing visual photos is a passive and user-friendly process that requires zero modification to existing network protocols. In contrast, acquiring instantaneous CSI for massive MTS arrays requires strict time-synchronization, overwhelming pilot overhead, and active multi-party protocol coordination between the base station, the user, and the MTS. Therefore, taking photos remains vastly more accessible than channel estimation in practical deployments (e.g., in sterile workshops, or public areas where base-station protocol modification is not authorized).

Next, we perform ray tracing for the 3D model to emulate the radio propagation. In particular, there are various material options for each object in the 3D model, including wood, metal, concrete, etc. The choice of material decides the electromagnetic properties that play a key role in ray tracing as specified in the ITU-R P.2040-2 recommendation. An open source software called \emph{Sionna} is adopted in our system for the ray tracing purpose, where the object material setting can be customized by editing the file titled ``radio\_material''. Note that the main task is to predict the distribution of blind spots, so any other digital-twin framework or high-quality prior knowledge of channels can be utilized here. Furthermore, we produce a grid map for the geographic area of the target network; ray tracing is performed with respect to the center of each grid cell. The above steps lead us to a gridded heatmap in terms of RSS, based on which the blind spots are sensed and clustered as discussed later.


\begin{figure*}[t]
    \centering
    \includegraphics[width=\linewidth]{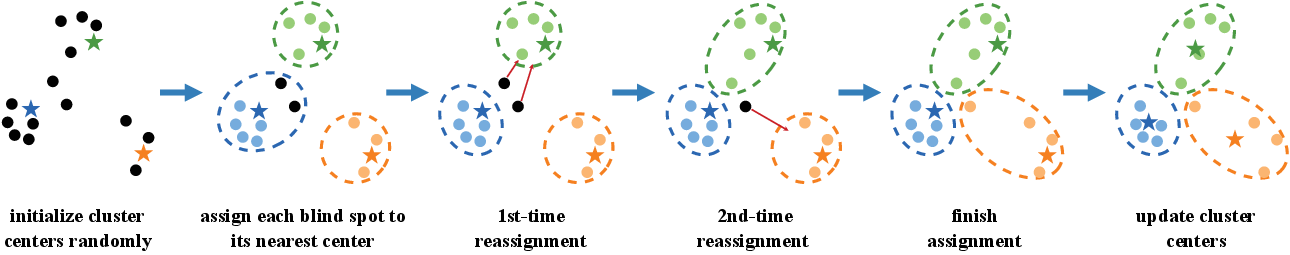}
    \caption{Procedure of a modified $K$-means clustering. In this example, we aim to divide $14$ blind spots into $3$ clusters, assuming that each cluster can contain $5$ blind spots at most. Note that two blind spots first choose the blue cluster, and later move to the green cluster because the blue cluster is full, but then the green cluster becomes full, so one of them further moves to the orange cluster.}
    \label{fig:modified_kmeans}
\end{figure*}

One may wonder why not just use ray tracing to predict channels so that the conventional CSI-based optimization method can follow immediately. The estimation error is the main reason. Ray tracing takes the 3D model as input, so it solely depends on the geographic topology of the radio environment. As a result, ray tracing is well suited for acquiring the large-scale fadings (as considered in \cite{2024GPMS,PMSat2023} for the oversimplified free space model), but the small-scale fadings are beyond its capability. 
The two questions Q1 and Q2, i.e., how many MTSs are needed and where the MTSs are placed, are expected to be solved in the long run, so it is reasonable to apply ray tracing here. Nevertheless, when it comes to the phase shift optimization in Q3 which is highly sensitive to the small-scale fadings, ray tracing is no longer sufficient. 


\subsection{Blind Spot Sensing and Clustering}

With ray tracing conducted for each grid cell, we can identify the blind spots based on the RSS prediction shown on the gridded heatmap. Specifically, a grid cell is labeled a blind spot if its predicted RSS (at its center) falls below a prescribed threshold. Clearly, if the threshold is raised, then more blind spots are identified, and the clusters become less scattered. As a result, the MTSs would be densely deployed. In practice, the choice of the blind spot threshold can be customized based on the modulation and coding scheme (MCS) according to the specific wireless service that we wish to accommodate.

We are now ready to answer Q1. If $M$ MTSs are deployed, then we shall divide the existing blind spots into $M$ clusters, and assign an MTS to each cluster. A common idea is to perform clustering by 
the classical K-means algorithm. However, since the reflection beam of MTS has a limited size, the number of blind spots that all the reflection beams from the same MTS can cover is limited as well. To address the above issue, we propose modifying the K-means algorithm as follows. Let us start by reviewing how the conventional K-means algorithm works: first, generate $M$ cluster centers randomly; second, assign each blind spot to the closest cluster center (in terms of the Euclidean distance); third, update each cluster center to minimize the within-cluster sum of the squared distances; repeat the previous two steps until the cluster centers converge. We now incorporate the cluster capacity constraint $C$ into the second step of the K-means algorithm: for each blind spot, if its closest cluster already contains $C$ blind spots, then try the second closest cluster center, and so forth; there must exist a cluster that the present blind spot can be assigned to because of the pigeonhole principle. This modified K-means algorithm for the blind spot clustering is illustrated in Fig. \ref{fig:modified_kmeans}.
We acknowledge that the individual components (3D modeling, K-means, ray tracing) are based on existing tools. However, we would like to highlight that the architectural synthesis of these tools into a fully automated, CSI-free, and closed-loop ``plug-and-play'' pipeline represents a significant system-level innovation for practical MTS deployment.

It remains to decide the number of clusters, namely the number of MTSs, denoted by $M$. We optimize $M$ in a greedy fashion. Initialize $M$ to the smallest possible number of MTSs to handle all the blind spots---which amounts to the ratio between the number of blind spots and the cluster capacity $C$. Following the idea of digital twin, we then virtually deploy these $M$ MTSs in the 3D model to see whether the existing blind spots would all disappear. Keep adding MTS until all the blind spots have been removed from the gridded heatmap of RSS. Note that the blind spots need to be re-sensed and re-clustered every after a new MTS has been added to the virtual 3D wireless environment.

Note also that we have not yet described where the $M$ MTSs should be placed. The MTSs are allowed to be placed anywhere within the feasible region. Recall that each MTS is associated with a cluster of blind spots, so the position of each MTS is closely related to where its blind spots are located. A simple geometric method is to put each MTS at a position so that the reflection path starting from the access point (AP) can pass through the cluster center of the  blind spots. In other words, we would place MTS where the angle of arrival from the AP equals the angle of departure toward the center of the cluster. However, as a subtle issue, the resulting position may not lie within the feasible region for the MTS deployment (e.g., people may not want an MTS to be placed in the middle of the room). To resolve this issue, we suggest the nearest projection onto the feasible region.

For larger indoor venues such as shopping malls or airports, the proposed algorithm can scale easily, since we just need
to take more photos (e.g., by unmanned aerial vehicles) and run the CSM algorithm separately on
each MTS. But we may encounter an aesthetic challenge: people possibly dislike seeing the cold metal panels of MTS distributed around. There are two solutions. One is to decorate MTSs into posters. The other is to deploy MTSs onto the ceiling, namely the ceiling-mounted MTS, which is quite suited for the high-ceiling venues like shopping malls and airports. In particular, if it is an mmWave system, we may need to deploy many more MTSs, but each MTS can be made smaller in size.



\section{Short-Run Configuration}
\label{sec:phase shifting}


Meta-atom phase shifting for the already deployed MTSs can be thought of as a fine-tuning adapted to the dynamic RF environment. As discussed earlier, the geographic features of the network extracted from the visual snapshots can only capture the distance-dependent attenuation, but the actual network is much more complicated because of the small-scale fadings and the user distribution. To capture these subtle aspects of RF environment, we propose exploiting the real-time RSRP data collected from the target receivers.

Again, we wish to avoid channel estimation. For ease of discussion, let us focus on the single-receiver case. The overall procedure of the phase shifting part of RFZero follows: (i) try out a sequence of random combinations of phase shifts across all the meta-atoms, each combination referred to as a sample; (ii) measure the RSS of the target receiver for each sample; (iii) compute the conditional sample mean of RSS with respect to every possible phase shift choice on each meta-atom; (iv) choose phase shift on each meta-atom to maximize the conditional sample mean. Here is a toy example to illustrate the above steps. Assume that two MTSs are deployed and that each MTS has only two meta-atoms. Let $\theta_{ij}$ be the phase shift of the $j$th meta-atom of the $i$th MTS; restrict the choice of phase shift to either $0$ or $\pi$, so each $\theta_{ij}$ takes on a value from the set $\{0,\pi\}$. We try out the following $6$ random samples of $(\theta_{11},\theta_{12},\theta_{21},\theta_{22})$: $(0,\pi,0,0)$, $(0,0,0,0)$, $(\pi,\pi,\pi,0)$, $(\pi,0,\pi,0)$, $(\pi,\pi,0,\pi)$, and $(0,0,\pi,\pi)$. The resulting RSS levels of these random samples are $2.8$, $1.0$, $1.5$, $3.3$, $0.3$, and $0.4$. Now let us focus on how $\theta_{11}$ is decided. First, we compute the CSM of RSS when $\theta_{11}$ is fixed at $0$, so the samples $\{1,2,6\}$ are used to obtain the CSM as $(2.8+1.0+0.4)/3=1.40$. Next, we average out the remaining samples $\{3,4,5\}$ to obtain the CSM as $(1.5+3.3+0.3)/3=1.70$ when $\theta_{11}$ is fixed at $\pi$. Comparing the above two CSM values, we set $\theta_{11}$ to $\pi$. The rest phase shifts can be decided similarly. The complete solution in this example is $\theta_{11}=\pi$, $\theta_{12}=0$, $\theta_{21}=\pi$, and $\theta_{22}=0$. The MTS reflection training can be performed \emph{ad hoc} so that random samples need not be consecutive, i.e., the samples are collected only when the base station transmit signals are available. Moreover, as shown in \cite{lai2024blind}, the non-i.i.d. random samples of phase shifts can possibly outperform the i.i.d. ones.

\begin{figure*}[t]
    \subfigure[Scenario I]{
    \includegraphics[width=0.3\linewidth]{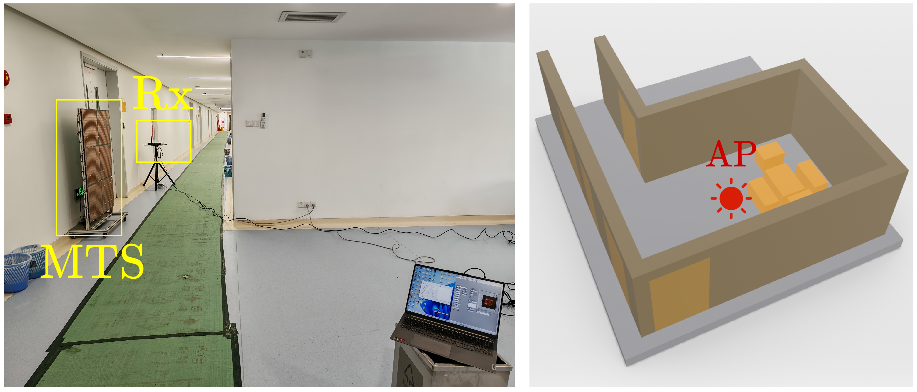}}
    \hfill
    \subfigure[Scenario II]{
    \includegraphics[width=0.3\linewidth]{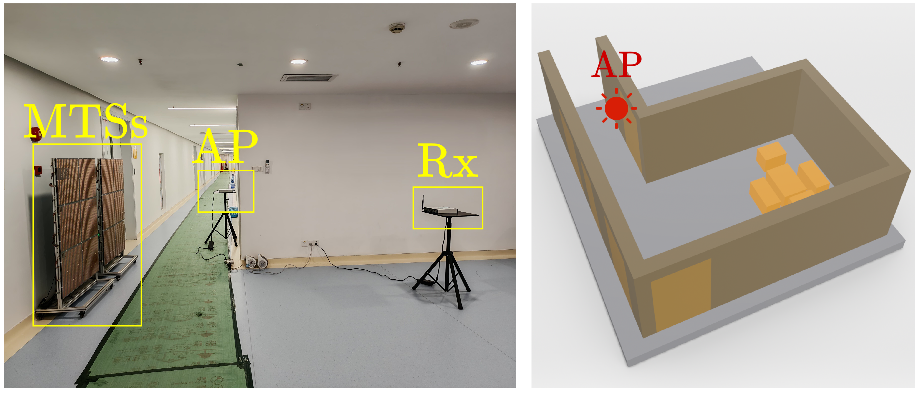}}
    \hfill
    \subfigure[Scenario III]{
    \includegraphics[width=0.3\linewidth]{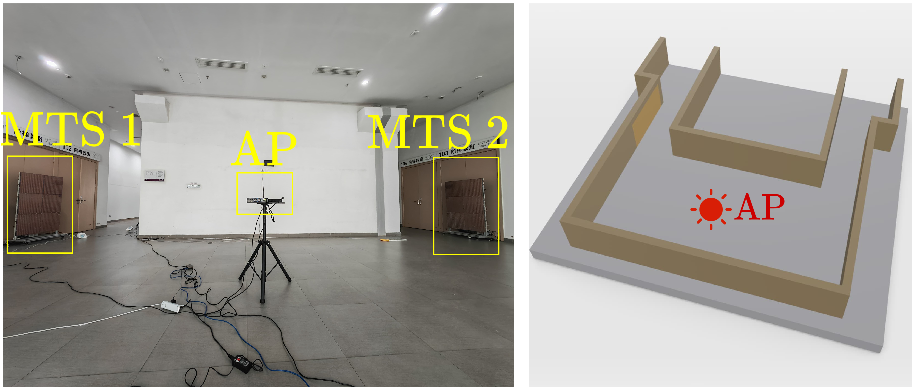}}
    \caption{The on-site photo (left) and the 3D model (right) for the three wireless network scenarios considered in the field tests.}
    \label{fig:scene_3D}
\end{figure*}

The above algorithm can be readily extended to account for multiple users. To start with, we consider the user-MTS association. This part can be customized according to the specific performance metrics adopted for the network. For instance, we may assign each user to its nearest MTS, or we may only consider those users with high priorities and assign each with multiple MTSs, etc. Let us focus on the phase shifting for a specific MTS after the user-MTS association has been determined. First, we require each user associated with this MTS to run the algorithm independently. Next, the phase shift of each meta-atom of this MTS is decided by majority voting among all users. When there are sufficiently many receivers, the extended CSM method is guaranteed to achieve a near-optimal solution \cite{xu2024blind}; the rationale is that the optimal solution should be favored by
most of the receivers. It is worthwhile to summarize what has been learned from data.
The modified K-means algorithm aims to learn the centroid of each blind spot cluster; the CSM algorithm aims to learn the phase distribution of the different reflected channels, thus aligning these reflected channels via phase shifting.

We now discuss the hardware impairment. If the phase shift error is fixed, then our method would not be impacted. If the phase shift error is random (and ergodic), then the convergence of our method would become slower. Moreover, the beam-point accuracy and sidelobe leakage can impact our solution to Q1 and Q2, e.g., we may need more MTSs than our algorithm predicts, and it can be suboptimal to position each MTS toward the centroid of the blind spot cluster.

We further define a hierarchical fallback mechanism in response to the environmental changes:
\begin{itemize}
    \item \emph{Level-1 Action (Phase Reconfiguration):} When the environment changes and a user's RSS drops, the system first triggers the short-term CSM algorithm to re-optimize the phase shifts of MTS.
    \item \emph{Level-2 Action (Visual Recapture):} If the worst-case RSS remains persistently below the blind-spot threshold after rerunning the CSM, this indicates that a large-scale blockage has fundamentally destroyed the line-of-sight propagation. Only when this threshold condition is triggered does our system alert the operator to capture new photos and reconstruct the 3D model for a new physical placement of MTS.
\end{itemize}
By introducing this RSS-threshold-driven trigger, the system avoids the overhead of unnecessary visual modeling while maintaining robust coverage.

\begin{figure*}[t]
    \centering
    \includegraphics[width=0.7\linewidth]{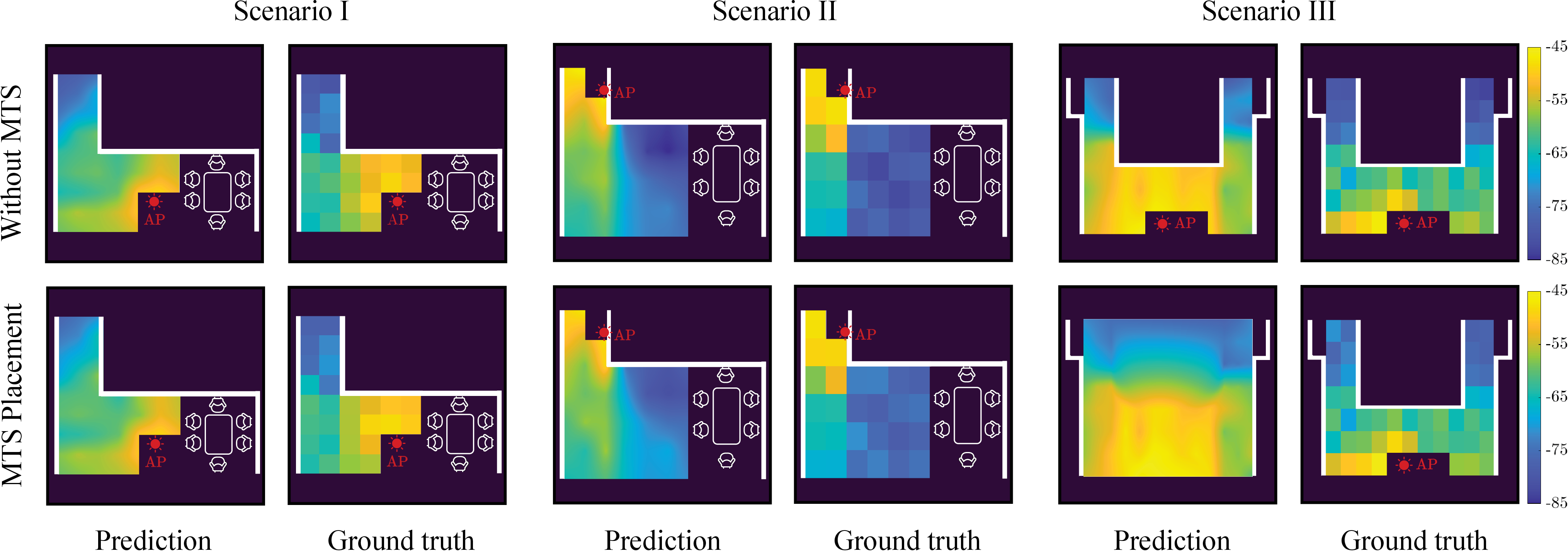}
    \caption{RSS (in dBm) prediction by RFZero vs. actual measurements.}
    \label{fig:simu_vs_field}
\end{figure*}


\section{Field Tests}
\label{sec:experiment}
\subsection{System Implementation}

\textbf{MTS prototype machines.} The MTS prototype machine used in our field tests is $1.5\text{ m}\times 0.9\text{ m}$ large, and consists of $21\times 14 = 294$ meta-atoms. The spacing between the centers of any two adjacent meta-atoms is $6$ cm. Equipped with a single PIN diode, each meta-atom can provide a binary phase shift choice: $0$ or $\pi$. The operating frequency band of MTS is $100$ MHz wide at $2.6$ GHz. In our case, the geographic area is partitioned into 1 m$\times$1 m square grid cells. Each MTS in our case can handle at most $6$ gridded blind spots.


    
        

\textbf{RF environment.} Three indoor RF environments as shown in Fig. \ref{fig:scene_3D} are considered here. Each of them is approximately $100\text{ m}^2$ large. Their 3D models based on photographs are also shown in Fig. \ref{fig:scene_3D}. The RSS threshold $\delta$ for the blind spot detection is set to $-78$ dBm \cite{shen2022a184} because $-78$ dBm is shown in \cite{shen2022a184} to be the lowest RSS for the sensor to detect the wake-up signal at the sub-6 GHz band.

\textbf{Communication system.}
We employ two NI-USRP X410 software-defined radios as the AP and the receiver (Rx) devices. A single omnidirectional transmit or receive antenna is deployed at each device. A narrow-band transmission is considered, in which the AP sends a signal at $2.6$ GHz with the bandwidth $125$ KHz. The sampling rate at the receiver side is $512$ KHz. Quadrature amplitude modulation (QAM) is used for modulation. All the performance metrics (e.g., RSS) are obtained from the platform NI LabVIEW v2021, measured every other 0.1 second.

\begin{figure*}[t]
    \centering

    \subfigure[RFZero]{
    \begin{minipage}{0.22\linewidth}
        \centering
        \includegraphics[width=3.8cm]{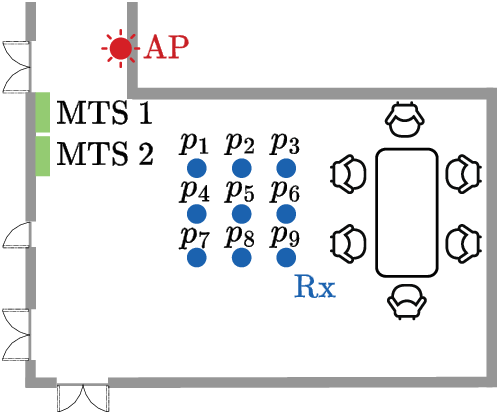}\vspace{5mm}\\
        \includegraphics[width=3.8cm]{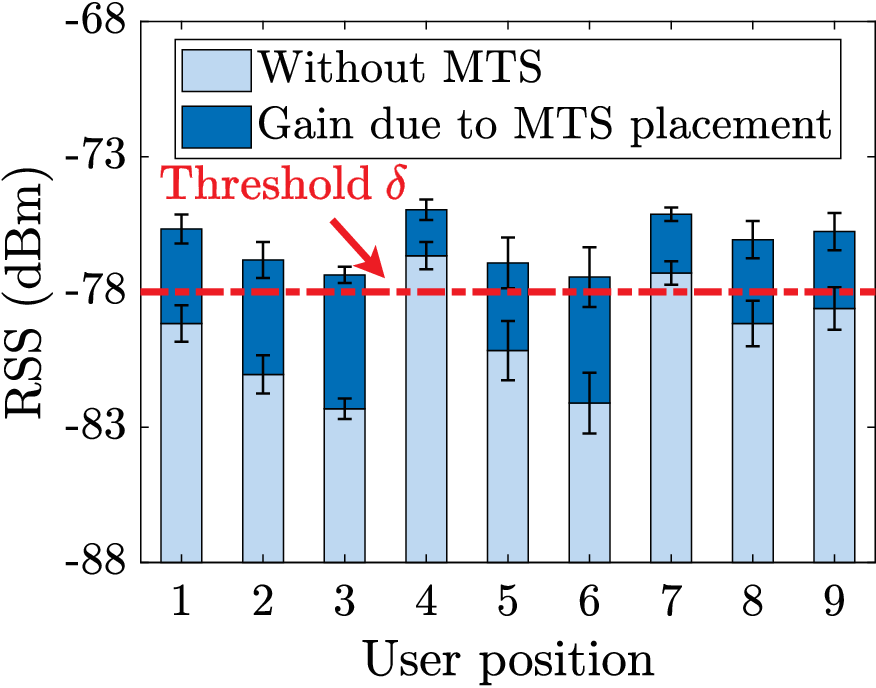}\vspace{2mm}
    \end{minipage}
    }
    \subfigure[AutoMS \cite{AutoMS2024}]{
    \begin{minipage}{0.22\linewidth}
        \centering
        \includegraphics[width=3.8cm]{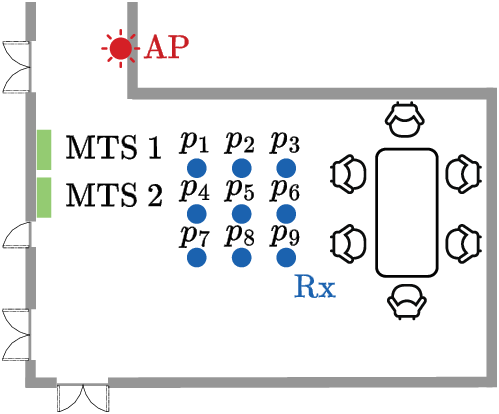}\vspace{5mm}\\
        \includegraphics[width=3.8cm]{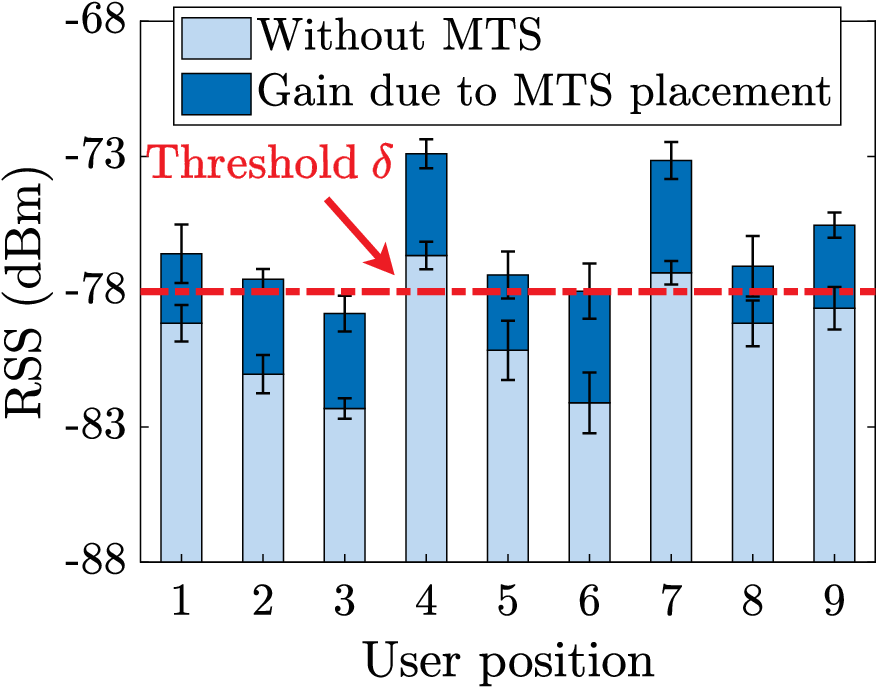}\vspace{2mm}
    \end{minipage}
    }
    \subfigure[Gradient-based placement \cite{van2023ris}]{
    \begin{minipage}{0.22\linewidth}
        \centering
        \includegraphics[width=3.8cm]{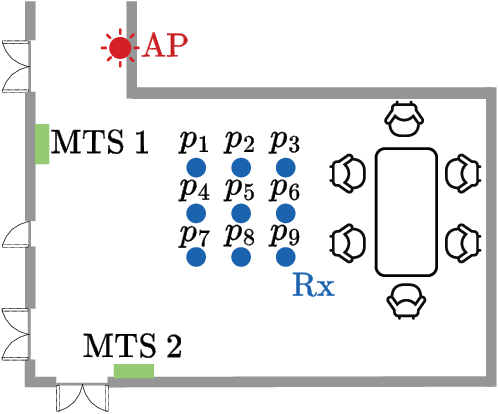}\vspace{5mm}\\
        \includegraphics[width=3.8cm]{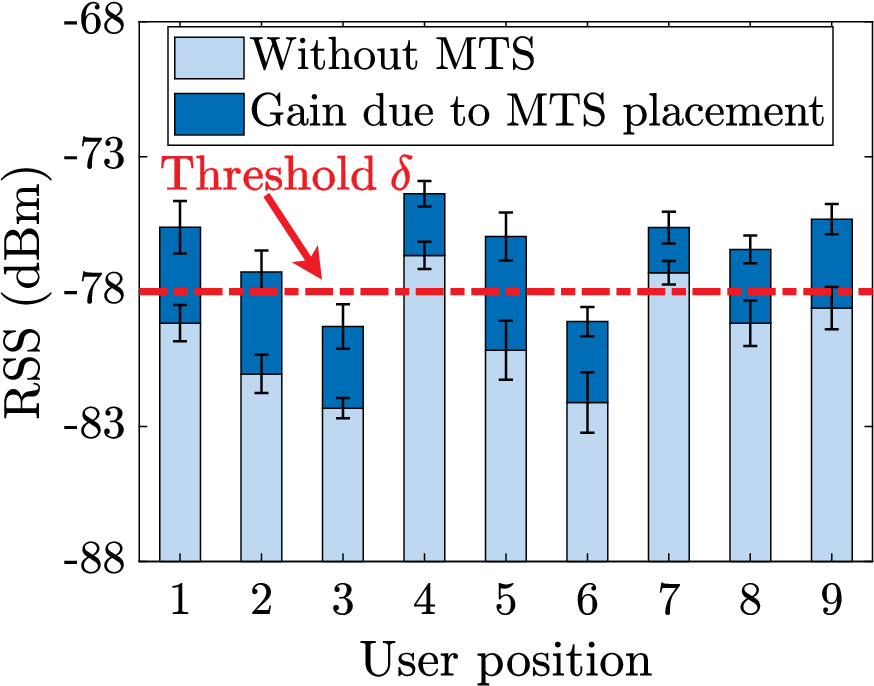}\vspace{2mm}
    \end{minipage}
    }
    \subfigure[Random placement]{
    \begin{minipage}{0.22\linewidth}
        \centering
        \includegraphics[width=3.8cm]{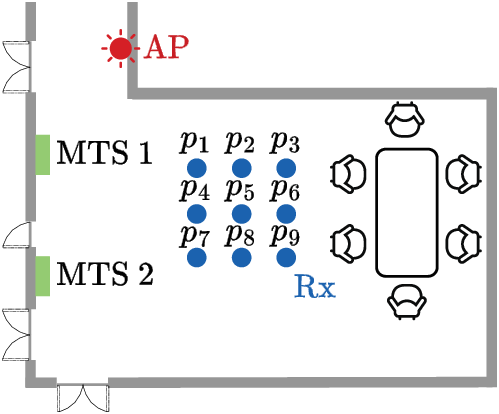}\vspace{5mm}\\
        \includegraphics[width=3.8cm]{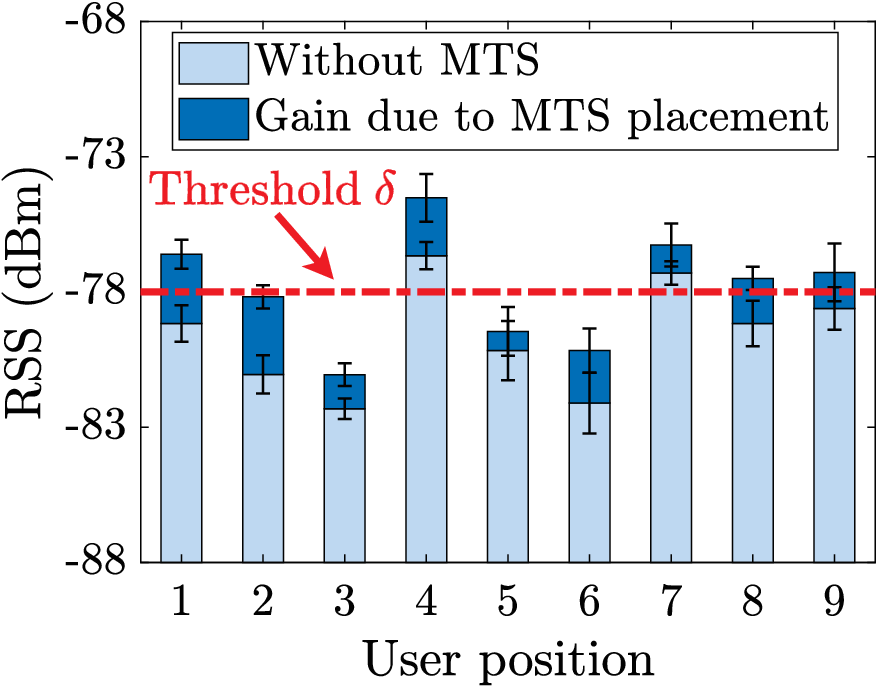}\vspace{2mm}
        \label{fig:random_place}
    \end{minipage}
    }
    
    \caption{Measured RSS at nine evaluated target spots under various MTS placement strategies for Scenario II.}
    \label{fig:placement}
\end{figure*}

\begin{figure*}[t]
\centering
\includegraphics[width=\linewidth]{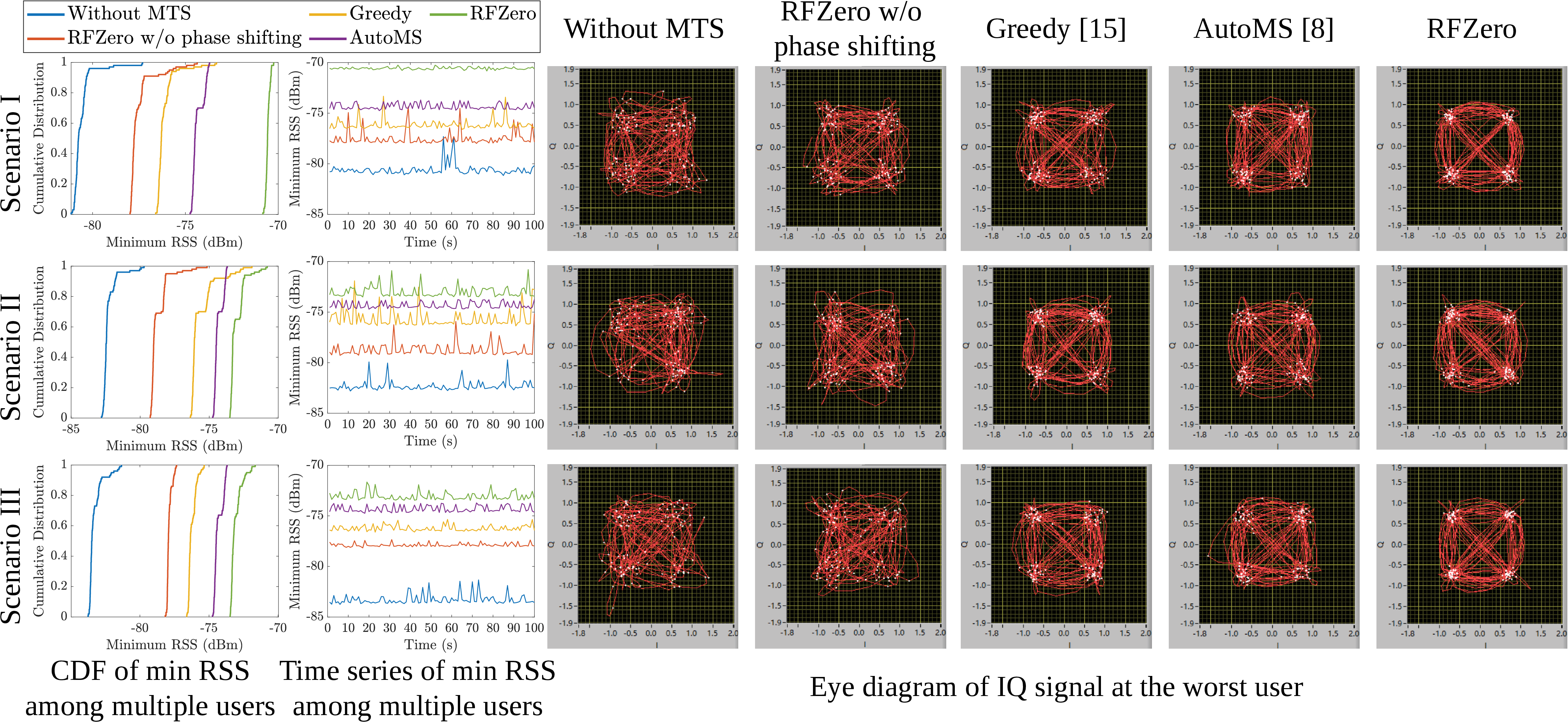}
\caption{Supplementary performance metrics for the different MTS placement and phase shifting methods.}
\label{fig:more_results}
\end{figure*}

\textbf{Benchmarks.} We compare the proposed RFZero with the following benchmarks: (i) \textit{Greedy \cite{chen2023seamless}:} for the fixed MTS placement by RFZero, it tries out a sequence of random choices of phase shifts and picks the best according to the min received signal power among users; (ii) \textit{AutoMS \cite{AutoMS2024}:} it optimizes the placement and phase shifts of MTS in an exhaustive manner. Notice that both RFZero and Greedy require $T$ random samples. For fairness, we let $T=1000$ by default for both methods.

\subsection{3D Modeling and RSS Prediction}
We now evaluate the 3D modeling and RSS prediction. As shown in Fig. \ref{fig:scene_3D}, all the three scenarios have been successfully reconstructed as the 3D model. Although many minor details are neglected by the 3D model, the key component, i.e., the geographic topology, has been well preserved, which is sufficient for the later RSS prediction.

Fig. \ref{fig:simu_vs_field} compares our RSS predictions with the ground truth. While high-resolution ray tracing yields nearly continuous predictions, the ground truth relies on discrete real-world measurements per grid cell. Nevertheless, our predictions align closely with the ground truth, particularly regarding the spatial distribution of blind spots. The average prediction errors for the three scenarios are approximately $3.4$ dB, $4.4$ dB, and $6.5$ dB, respectively.

\subsection{Blind Spot Elimination}

We now evaluate the performance gain due to the MTS placement. As shown in Fig. \ref{fig:placement}, we compare the RSS at nine user positions before and after the MTS deployment. We aim to maximize the worst position so that all the positions are above the blind spot threshold $\delta$. Aside from the MTS placement by RFZero, we consider the placements produced by AutoMS \cite{AutoMS2024} and the gradient-based placement optimizer \cite{van2023ris}. Moreover, we randomly select a placement as the benchmark as shown in Fig. \ref{fig:random_place}. Fig. \ref{fig:placement} summarizes the performance of the different placement algorithms with phase shifts fixed at zero for each meta-atom. Throughout this experiment, we let $\delta=-78$ dBm. Observe that only the proposed RFZero is capable of eliminating all the blind spots, whereas AutoMS would lead to position 3 being a blind spot and the gradient-based optimizer still leaves two severe blind spots. We further compare the RSS gain between the blind spots and non-blind spots. It can be seen that RFZero provides much higher RSS gain for those blind spots; this makes sense since the blind spot elimination is determined by the worst user. In contrast, AutoMS is less sensitive to the blind spot, providing roughly the same gain to all positions. The average RSS gains across all blind spots achieved by RFZero and AutoMS are $3.8$ dB and $3.1$ dB, respectively. 


\subsection{RSS Enhancement by Phase Shifting}

Finally, let us look at how much gain the deployed MTSs can bring to the target users. Consider this baseline: place the MTSs by RFZero but do not optimize phase shifts. Two other benchmarks are from the existing literature: Greedy \cite{chen2023seamless} and AutoMS \cite{AutoMS2024}. (Since Greedy only accounts for phase shifting, we use RFZero to decide the MTS placement for it.) 
Fig. \ref{fig:more_results} compares the different phase shifting algorithms. In the first column of Fig. \ref{fig:more_results}, we compare the cumulative distribution function (CDF) of the min RSS among users, assuming that one user is located at each spot. Observe that the CDF curve of RFZero is on the right of the rest CDF curves, thus RFZero has the best performance in the low-RSS regime. Moreover, from the second column of Fig. \ref{fig:more_results}, we notice that the RSS of RFZero is stable over time in scenarios I and III; it has stronger fluctuations in scenario II as other methods. Last, we draw the eye diagram of IQ signals at the lowest RSS position. Observe that the eyes of RFZero ``open'' more widely than those of other algorithms, so RFZero yields a better RF environment with a lower intersymbol interference (ISI) level.


\section{Conclusion}
\label{sec:Conclusion}
This paper proposes RFZero, a zero-CSI paradigm for deploying and operating MTSs to eliminate wireless blind spots. RFZero decouples the problem into two timescales: leveraging visual photos for long-term MTS placement, and utilizing RSRP feedback for short-term phase coordination. By circumventing explicit channel estimation, RFZero enables truly plug-and-play deployment. Field tests at $2.6$ GHz confirm that this strategy not only completely eliminates network blind spots but also delivers an approximate $10$ dB SNR improvement throughout the wireless environment.


\section{Acknowledgements}

This work was supported
in part by Guangdong Major Project of Basic and Applied
Basic Research (No. 2023B0303000001) and in part by NSFC No. 92167202.

\bibliographystyle{IEEEtran}     
\bibliography{IEEEabrv,ref}

@STRING{IEEE_J_COM        = "{IEEE} Trans. Commun."}

@STRING{IEEE_J_WCOM       = "{IEEE} Trans. Wireless Commun."}

@STRING{IEEE_J_WCOML      = "{IEEE} Wireless Commun. Lett."}

@STRING{IEEE_J_MTT        = "{IEEE} Trans. Microw. Theory Techn."}

@STRING{IEEE_ICASSP        = "Proc. {IEEE} Int. Conf. Acoust., Speech, Signal Process. (ICASSP)"}

@inproceedings{Arun2020RFocus,
author = {V. Arun and H. Balakrishnan},
title = "{RFocus:} Beamforming Using Thousands of Passive Antennas",
booktitle = "{USENIX} Symp. Netw. Sys. Design Implementation ({NSDI})",
year = 2020,
isbn = {978-1-939133-13-7},
pages = {1047-1061},
month = feb
}

@article{pei2021ris,
  title="{RIS}-aided wireless communications: Prototyping, adaptive beamforming, and indoor/outdoor field trials",
  author="Pei, Xilong and Yin, Haifan and Tan, Li and Cao, Lin and Li, Zhanpeng and Wang, Kai and Zhang, Kun and Bj{\"o}rnson, Emil",
  journal=IEEE_J_COM,
  volume=69,
  number=12,
  pages={8627--8640},
  year=2021,
  month=dec
}

@inproceedings{2024GPMS,
author = {Wang, Yezhou and Pan, Hao and Qiu, Lili and Zhong, Linghui and Liu, Jiting and Ma, Ruichun and Chen, Yi-Chao and Xue, Guangtao and Ren, Ju},
title = {{GPMS}: Enabling Indoor {GNSS} Positioning using Passive Metasurfaces},
year = {2024},
booktitle = {Proc. 30th Annu. Int. Conf. Mobile Comput., Netw. (MobiCom)},
pages = {1424-1438},
month=may
}

@inproceedings {RFBouncer2023,
author = {Xinyi Li and Chao Feng and Xiaojing Wang and Yangfan Zhang and Yaxiong Xie and Xiaojiang Chen},
title = {{RF-Bouncer}: A Programmable Dual-band Metasurface for Sub-6 Wireless Networks},
booktitle = "{USENIX} Symp. Netw. Sys. Design Implementation ({NSDI})",
year = {2023},
pages = {389--404},
month=apr
}

@inproceedings{PMSat2023,
author = {Pan, Hao and Qiu, Lili and Ouyang, Bei and Zheng, Shicheng and Zhang, Yongzhao and Chen, Yi-Chao and Xue, Guangtao},
title = {{PMSat}: Optimizing Passive Metasurface for Low Earth Orbit Satellite Communication},
year = {2023},
booktitle = {Proc. 29th Annu. Int. Conf. Mobile Comput., Netw. (MobiCom)},
pages = {1-15},
month=oct
}

@inproceedings{chen2023seamless,
author = {Chen, Lili and Yu, Bozhong and Ren, Ju and Gummeson, Jeremy and Zhang, Yaoxue},
title = {Towards Seamless Wireless Link Connection},
year = {2023},
booktitle = {Proc. 21th Annu. Int. Conf. Mobile Syst., Appl., Services (MobiSys)},
pages = {137-149},
month=jun
}

@inproceedings{AutoMS2024,
author = {Ma, Ruichun and Zheng, Shicheng and Pan, Hao and Qiu, Lili and Chen, Xingyu and Liu, Liangyu and Liu, Yihong and Hu, Wenjun and Ren, Ju},
title = {{AutoMS}: Automated Service for mm{W}ave Coverage Optimization using Low-cost Metasurfaces},
year = {2024},
booktitle = {Proc. 30th Annu. Int. Conf. Mobile Comput., Netw. (MobiCom)},
pages = {62-76},
month=may
}

@inproceedings{SmartShell2023,
author = {Zhong, Linling and Ouyang, Mingwei and Zhu, Fengyuan and Jin, Meng and Wang, Xinbing and Guan, Xinping and Zhou, Chenghu and Tian, Xiaohua},
title = {{SmartShell}: A Near-Field Reflective Surface Enhancing {RSS}},
year = {2023},
booktitle = {Proc. 21th Annu. Int. Conf. Mobile Syst., Appl., Services (MobiSys)},
pages = {124-136},
month=jun
}

@Unpublished{roomplan,
  title={RoomPlan API.},
  author={Apple Developer.},
  note = "https://developer.apple.com/ augmented-reality/roomplan",
  year=2023
}

@article{efrem2023joint,
  title={Joint {IRS} location and size optimization in multi-{IRS} aided two-way full-duplex communication systems},
  author={Efrem, Christos N and Krikidis, Ioannis},
  journal=IEEE_J_WCOM,
  volume={22},
  number={10},
  pages={6518--6533},
  year={2023},
  month=oct
}

@software{sionna,
 title = {Sionna},
 author = {Hoydis, Jakob and Cammerer, Sebastian and {Ait Aoudia}, Fayçal and Nimier-David, Merlin and Maggi, Lorenzo and Marcus, Guillermo and Vem, Avinash and Keller, Alexander},
 note = {https://nvlabs.github.io/sionna/},
 year = {2022},
 version = {0.19.2},
 organization = {NVIDIA}
}

@article{mu2021capacity,
  title={Capacity and optimal resource allocation for {IRS}-assisted multi-user communication systems},
  author={Mu, Xidong and Liu, Yuanwei and Guo, Li and Lin, Jiaru and Al-Dhahir, Naofal},
  journal=IEEE_J_COM,
  volume={69},
  number={6},
  pages={3771--3786},
  year={2021},
  month=jun
}

@article{shen2022a184,
  title={A 184-{nW}, -78.3-{dBm} sensitivity antenna-coupled supply, temperature, and interference-robust wake-up receiver at 4.9 {GHz}},
  author={Shen, Xiaochuan and Duvvuri, Divya and Bassirian, Pouyan and Bishop, Henry L and Liu, Xinjian and Dissanayake, Anjana and Zhang, Yaobin and Blalock, Travis N and Calhoun, Benton H and Bowers, Steven M},
  journal=IEEE_J_MTT,
  volume={70},
  number={1},
  pages={744--757},
  year={2022},
  month=jan
}

@article{xu2024blind,
  title={Blind Beamforming for Coverage Enhancement with Intelligent Reflecting Surface},
  author={Xu, Fan and Yao, Jiawei and Lai, Wenhai and Shen, Kaiming and Li, Xin and Chen, Xin and Luo, Zhi-Quan},
  journal=IEEE_J_WCOM,
  volume=23,
  number=11,
  pages={15736--15752},
  year={2024},
  month=nov
}

@article{zheng2019intelligent,
  title={Intelligent reflecting surface-enhanced {OFDM}: Channel estimation and reflection optimization},
  author={Zheng, Beixiong and Zhang, Rui},
  journal=IEEE_J_WCOML,
  volume={9},
  number={4},
  pages={518--522},
  year=2020,
  month=apr
}

@inproceedings{lai2024blind,
  title={Blind beamforming for intelligent reflecting surface: A reinforcement learning approach},
  author={Lai, Wenhai and Shen, Kaiming},
  booktitle=IEEE_ICASSP,
  pages={8956--8960},
  year={2024},
  month=may
}

@article{van2023ris,
  title={{RIS}-assisted wireless communications: Long-term versus short-term phase shift designs},
  author={Van Chien, Trinh and Tu, Lam-Thanh and Khalid, Waqas and Yu, Heejung and Chatzinotas, Symeon and Di Renzo, Marco},
  journal=IEEE_J_COM,
  volume={72},
  number={2},
  pages={1175--1190},
  year={2024},
  month=feb
}

\begin{IEEEbiographynophoto}{Wenhai Lai} 
(wenhailai@dlut.edu.cn) was a Ph.D. student with The Chinese University of Hong Kong (Shenzhen). He is now an Assistant Professor with the School of General Education, Dalian University of Technology. His research interests include metasurface, optimization, and machine learning. 
\end{IEEEbiographynophoto}

\begin{IEEEbiographynophoto}{Mingxiao Li} 
(mingxiaoli@link.cuhk.edu.cn) is currently pursuing the Ph.D. degree at The Chinese University of Hong Kong (Shenzhen). His research interests include communications and machine learning.
\end{IEEEbiographynophoto}

\begin{IEEEbiographynophoto}{Kaiming Shen}
(shenkaiming@cuhk.edu.cn) is an Assistant Professor with the School of Science and Engineering, The Chinese University of Hong Kong (Shenzhen). His research interests include information theory, signal processing, and machine learning. 
\end{IEEEbiographynophoto}

\begin{IEEEbiographynophoto}{Liyao Xiang}
(xiangliyao08@sjtu.edu.cn) is currently an Associate Professor at Shanghai Jiao Tong University. Her
research interests include security and privacy, privacy analysis in data mining, and mobile computing.
\end{IEEEbiographynophoto}

\begin{IEEEbiographynophoto}{Zhi-Quan Luo}
(luozq@cuhk.edu.cn) is currently the Vice President (Academic) of The Chinese University of Hong Kong, Shenzhen, where he has been a Professor since 2014. He is concurrently the Director of the Shenzhen Research Institute of Big Data. His research interests lie in the area of optimization, big data, signal processing and digital communication.
\end{IEEEbiographynophoto}

\end{document}